\documentclass[twocolumn,aps,showpacs,preprintnumbers,lettersize]{revtex4}
\usepackage{amsfonts}
\usepackage{amsmath}
\usepackage{amssymb}
\usepackage{graphicx}

\begin{document}

\title{Interaction-Induced Adiabatic Cooling for Antiferromagnetism in
Optical Lattices}
\author{A.-M. Dar\'{e}, L. Raymond, G. Albinet}
\affiliation{L2MP, B\^{a}timent IRPHE, 49 rue Joliot Curie BP 146, Universit\'{e} de
Provence, 13384 Marseille, Cedex 13, France }
\author{A.-M. S. Tremblay}
\affiliation{D\'{e}partement de physique and Regroupement qu\'{e}b\'{e}cois sur les mat%
\'{e}riaux de pointe, Universit\'{e} de Sherbrooke, Sherbrooke, Qu\'{e}bec,
J1K 2R1, Canada}
\date{\today }

\begin{abstract}
In the experimental context of cold-fermion optical lattices, we discuss the possibilities to approach the pseudogap or ordered phases by manipulating the scattering length or the strength of the laser-induced lattice potential.
Using the Two-Particle Self-Consistent Approach as well as Quantum Monte
Carlo simulations, we provide isentropic curves for the two- and
three-dimensional Hubbard models at half-filling. These quantitative results
are important for practical attempts to reach the ordered antiferromagnetic
phase in experiments on optical lattices of two-component fermions. We find
that adiabatically turning on the interaction in two dimensions to cool
the system is not very effective. In three dimensions, adiabatic cooling to
the antiferromagnetic phase can be achieved in such a manner although the
cooling efficiency is not as high as initially suggested by Dynamical
Mean-Field Theory. Adiabatic cooling by turning off the repulsion
beginning at strong coupling is possible in certain cases.

\end{abstract}

\maketitle

\section{Introduction}

One of the most exciting possibilities opened by research on cold atoms in
optical traps is to study in a controlled manner model Hamiltonians of
interest to condensed matter physics. For example, high on the list of
questions that can in principle be answered by these model systems is
whether high-temperature superconductivity can be explained by the
two-dimensional Hubbard model away from half-filling. \cite%
{Jaksch:2004,Hofstetter:2002} As a first step towards achieving this goal,
the antiferromagnetic phase expected at half-filling offers an easier target
state that occurs at higher temperature in the phase diagram of cuprate
superconductors. \cite{Damascelli:2003,Lee:2004}

In optical lattices, the two spin species occurring in the Hubbard model are
mimicked by atoms in two different hyperfine states. The cooling of these
atomic gases necessary to observe ordered states has been discussed before.
\cite{Blackie:2005,Hofstetter:2002} It has been recently pointed out
however that there is an additional mechanism, \cite{Werner:2006} akin to
Pomeranchuck cooling in liquid Helium 3, that is available to help in
achieving the temperatures where antiferromagnetism can be observed. In this
mechanism, the temperature can be lowered by turning on at constant entropy
what amounts to interactions in the Hubbard model
(see Ref.(\onlinecite{Georges:2007}) for a review). The original calculations
for this effect were done for the Hubbard model using Dynamical Mean-Field
Theory (DMFT). \cite{Werner:2006} While it is expected that this approach
will give qualitatively correct results, accurate predictions are necessary
to achieve the practical implementation of this cooling scheme. In the
present paper, we present such quantitative predictions for the isentropic
curves of both the two- and three-dimensional Hubbard models.
We display the results in the usual units for the Hubbard model, but also in the
conventional units used in the context of cold atom physics.

Solving the
problem for both two- and three-dimensional lattices fulfills several
purposes. First, the two-dimensional case is interesting in its own right
even if long-range order cannot be achieved at finite temperature in
strictly two dimensions (because of the Mermin-Wagner-Hohenberg theorem).
Indeed, high-temperature parent antiferromagnetic compounds have a strong
two-dimensional character. In addition, even though long-range order cannot
be achieved, there is a two-dimensional regime with very strong
antiferromagnetic fluctuations that is interesting in itself. In this
regime, a pseudogap appears in the single-particle spectral weight that is
caused at weak coupling by antiferromagnetic fluctuations that have a
correlation length larger than the single-particle de Broglie wavelength.
\cite{Vilk:1997,Tremblay:2006} At strong coupling, the pseudogap appears
well before the long antiferromagnetic correlation lengths occur. \cite%
{Kyung:2006} Also, considering both two- and three dimensions sheds
additional light on the mechanism for cooling.

The Hubbard model is defined in second quantization by,%
\begin{equation}
H=-\sum_{i,j,\sigma }t_{ij}c_{i\sigma }^{\dagger }c_{j\sigma
}+U\sum_{i}n_{i\uparrow }n_{i\downarrow }
\end{equation}%
where $c_{i\sigma }^{\dagger }$ ($c_{i\sigma }$) are creation and
annihilation operators for electrons of spin $\sigma $, $n_{i\sigma
}=c_{i\sigma }^{\dagger }c_{i\sigma }$ is the density of spin $\sigma $
electrons, $t_{ij}=t_{ji}^{\ast }$ is the hopping amplitude, and $U$ is the
on-site repulsion obtained from matrix elements of the contact interaction
between atoms in the basis of Wannier states of the optical lattice.\cite%
{Jaksch:2004,Georges:2007} We restrict ourselves to the case where only nearest-neighbor
hopping $t\ $coming from tunneling between potential minima is relevant.
In keeping with common practice, $t$ will be the energy unit, unless explicitly
stated.

Let us recall the physics of the cooling mechanism proposed in Ref.\ \cite%
{Werner:2006}. If we denote by $f$ the free-energy per lattice site and $s$
the corresponding entropy, then $s=-\left( \partial f/\partial T\right) _{U}$
and $d=\left( \partial f/\partial U\right) _{T}$ where $d=\langle
n_{\uparrow }n_{\downarrow }\rangle $ is the double-occupancy. The chemical
potential is kept constant in all partial derivatives without further
notice. In the particle-hole symmetric case at half-filling the density is
also constant at constant chemical potential. We thus have a Maxwell relation%
\begin{equation}
\left( \frac{\partial s}{\partial U}\right) _{T}=-\left( \frac{\partial d}{%
\partial T}\right) _{U}.  \label{Maxwell}
\end{equation}%
Following Ref.\ \cite{Werner:2006}, the shape of the isentropic curves $%
s\left( T_{i}\left( U\right) ,U\right) =cst$ can be deduced by taking a
derivative of the last equation and using the Maxwell relation Eq.(\ref%
{Maxwell})%
\begin{equation}
c\left( T_{i}\right) \left( \frac{\partial T_{i}}{\partial U}\right)
_{s}=T_{i}\left( \frac{\partial d}{\partial T}\right) _{U}
\end{equation}%
where $c\left( T_{i}\right) =T\left( \partial s/\partial T\right) _{U}$ is
the specific heat. If $\left( \partial d/\partial T\right) _{U}$ is negative
at small $U,$ then $\left( \partial T_{i}/\partial U\right) _{s}$ will be
negative and hence it will be possible to lower the temperature at constant
entropy by increasing $U.$ Generally, double occupancy increases with
temperature so $\left( \partial d/\partial T\right) _{U}$ is positive, but
it does happen that $\left( \partial d/\partial T\right) _{U}$ is negative,
leading to a minimum at some temperature. This result may seem
counterintuitive. Indeed, at strong coupling, namely for interaction
strength much larger than the bandwidth, such a phenomenon does not occur.
Double occupancy is already minimum at zero temperature. It only increases
with increasing temperature. At weak coupling however, when the temperature
is large enough that it allows states to be occupied over a large fraction
of the whole Brillouin zone, the electrons may become more localized than at
lower temperature. An alternate way to understand this minimum is to notice
that it occurs when the thermal de Broglie wavelength is of the order of the
lattice spacing: \cite{Lemay:2000} At larger temperatures, double-occupancy
increases because of thermal excitation while at lower temperature the
plane-wave nature of the states becomes more apparent and double occupancy
also increases. The minimum in $\left( \partial d/\partial T\right) _{U}$
has been observed in DMFT \cite{Georges:1996,Werner:2006} and also very
weakly in Quantum Monte Carlo (QMC) simulations of the two-dimensional model
at $U=4t$ (see Figure 3 of Ref. \cite{Paiva:2001}) while in the Two-Particle
Self-Consistent (TPSC) approach \cite{Vilk:1997,Allen:2003,Tremblay:2006}
that we employ along with QMC, very shallow minima are observed in three
dimensions and are barely observable in two dimensions depending on the
value of $U.$ \cite{Dare:2000,Lemay:2000,Kyung:2003a,Roy:2006} As we shall
see, in three dimensions a minimum in $\left( \partial d/\partial T\right)
_{U}$ is also predicted by second-order perturbation theory. \cite{Lemay:2000}%

In the next section we discuss the two methods that we use, emphasizing the
points that are specific to this problem. Then we present the results for
the constant entropy curves in two and three dimensions and conclude. Two
appendices present calculational details for the entropy curves in limiting
cases.

\section{Methodology{}}

In this section we give methodological details that are specific to this
work, referring to the literature for more detailed explanations of the QMC
and TPSC approaches.

\subsection{Quantum Monte Carlo simulations}

In two dimensions we perform QMC simulations following the
Blankenbecler-Sugar-Scalapino-Hirsch (determinantal) algorithm. \cite%
{Blankenbecler:1981} The standard formula to obtain the entropy consists in
integrating the specific heat. However, the evaluation of the latter
quantity involves a numerical derivative. To avoid differentiating data that
contains statistical uncertainty, we follow Ref. \cite{Werner:2006} and
perform an integration by parts to compute the entropy from the energy
density $e$%
\begin{equation}
s\left( \beta ,U\right) =\ln 4+\beta e\left( \beta ,U\right)
-\int_{0}^{\beta }e\left( \beta ^{\prime },U\right) d\beta ^{\prime }
\label{s_integration}
\end{equation}%
with $\beta =1/T$ in units where the Boltzmann constant equals unity. This uses
the fact that the entropy at infinite temperature is known exactly. The
integral is calculated from the trapezoidal rule on a grid of about twenty
points spread on a logarithmic scale that extends from $\beta =0$ to $\beta $
of order $5$ depending on the cases. Each data point is obtained by up to $%
15\times 10^{6}$ measurements for the $4\times 4$ lattices and $10^{6}$
measurements for $8\times 8$. By comparing with the known result at $U=0,$
we deduce that the error on the integral is of order $2$ to $3\%$ at most at
the lowest temperatures. At large values of $U,$ the systematic error due to
the discretization of the imaginary time can be quite large. We checked with
$U=14$, $\Delta \tau =1/10$, $1/20$ and $1/40$ (in units $t=1$ which we
adopt from now on) that $\Delta \tau =1/10$ and $1/20$ suffice for an
accurate $\Delta \tau \rightarrow 0$ extrapolation.

Size dependence becomes important at low temperature. These effects can be
estimated from the $U=0$ case. \cite{Paiva:2001} The usual formula for the
entropy%
\begin{equation}
s\left( T,U=0\right) =-\frac{2}{N}\sum_{\mathbf{k}}\left( f\ln f+\left(
1-f\right) \ln \left( 1-f\right) \right)  \label{s_U=0}
\end{equation}%
with $N=L\times L$ (and $L$ even) the number of sites and $f$ the
Fermi-Dirac distribution, leads to a residual entropy at $T=0$ given by%
\begin{equation}
s\left( 0,0\right) =\frac{2L-2}{L^{2}}\ln 4
\end{equation}%
which does vanish for $L\rightarrow \infty $ but which gives important
contributions for finite $L.$ For example, for $L=4,$ we have $s=0.52,$
compared with $\ln 4=1.386$ at $T=\infty $. This entropy is easily
understood by counting the number of ways to populate the states that are
right at the Fermi surface of the finite lattice in the half-filled Hubbard
model. \cite{Note-Degenerescence} At $T=0.3,$ one can check that the
relative error between the $4\times 4$ lattice and the infinite lattice is
about $30\%$ while for the $8\times 8$ lattice it is about $5\%$. At $T=0.5,$
the finite size error for the $8\times 8$ lattice is negligible while it is
about $5\%$ for the $4\times 4$ lattice. Since in this work we concentrate
on high-temperature results, this will in general not be a problem in QMC.
The TPSC calculations can be performed in the infinite-size limit and for a
finite-size lattice.

\subsection{Two-Particle Self-Consistent Approach}

The TPSC approach has been extensively checked against QMC approaches in
both two \cite{Vilk:1997,Tremblay:2006} and three \cite{Dare:2000}
dimensions. It is accurate from weak to intermediate coupling, in other
words for $U$ less than about $3/4$ of the bandwidth, namely $U=6$ in $d=2$.
The double occupancy is one of the most accurate quantities that can be
calculated at the first step of the TPSC calculation using sum rules. Hence,
we can compute the entropy directly by integrating the Maxwell relation Eq.(%
\ref{Maxwell}) in the thermodynamic limit, or for a finite-size system, using the known value of the
entropy at $U=0,$ Eq.(\ref{s_U=0}), to determine the integration constant.
Earlier results obtained with TPSC for the double occupancy may be found for
example in Refs. \cite{Vilk:1997,Dare:2000,Kyung:2003a}. Issues of
thermodynamic consistency have been discussed in Ref. \cite{Roy:2002}.

\begin{figure}[tbp]
\includegraphics[width=8.5cm]{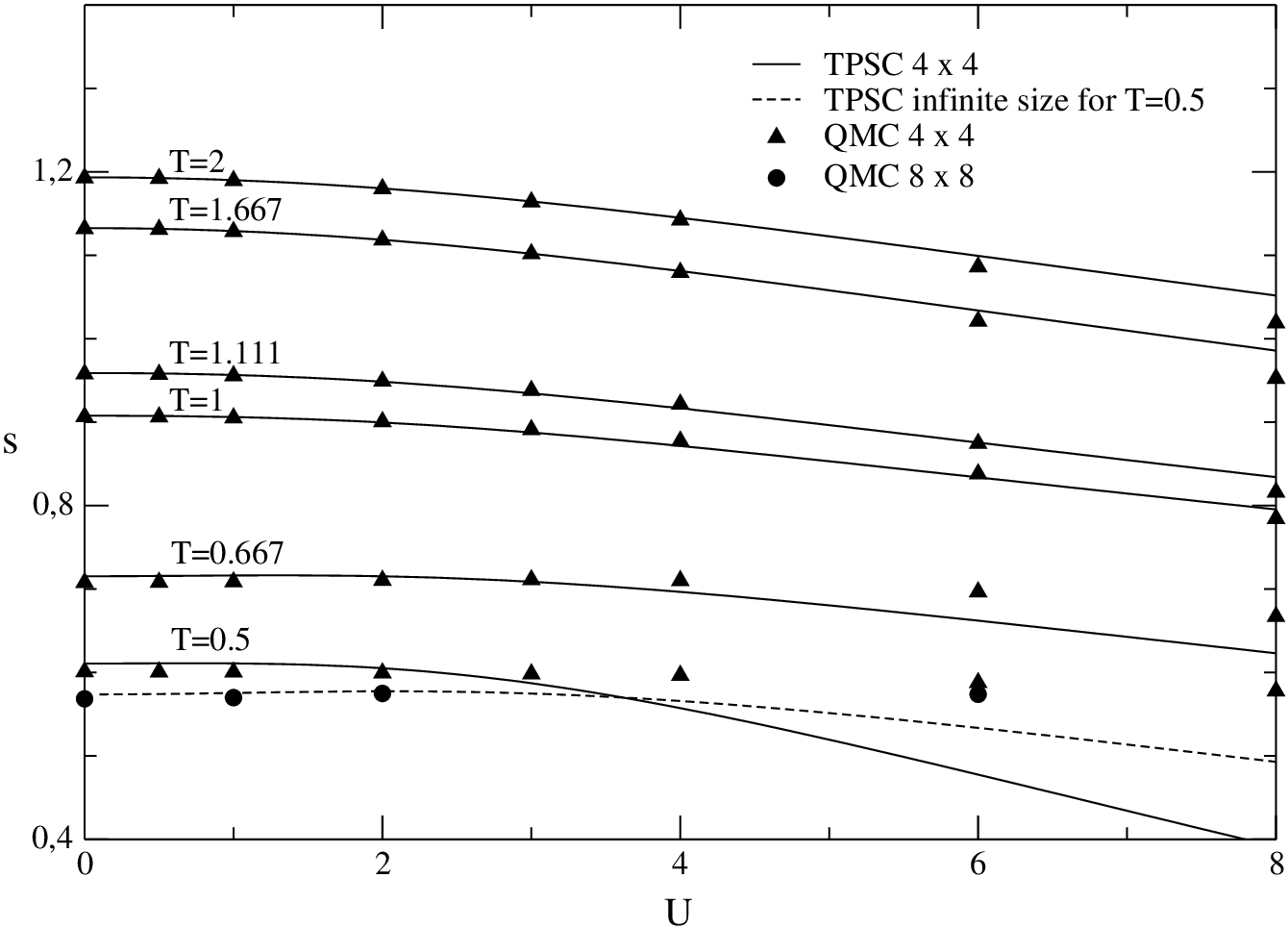}
\caption{Comparison of TPSC and QMC results for the entropy as a function of
$U$ for different temperatures. Solid lines are for TPSC $4\times 4,$
triangles for QMC $4\times 4.$ The dashed line is the TPSC result for the
infinite-size lattice limit for $T=0.5$. Also shown by filled circles for this
temperature, are the results for QMC $8\times 8.$ }
\label{TPSCvsQMC}
\end{figure}

\qquad Fig. \ref{TPSCvsQMC} compares the entropy obtained with TPSC (solid
and dashed lines) and with QMC (symbols) as a function of $U$ in $d=2$ for
different temperatures. The QMC calculations are for a $4\times 4$ system
except for $T=0.5$ where we also show results for $8\times 8.$ The TPSC
calculations are presented for both $4\times 4$ (solid line) and infinite
size limit (dashed line for $T=0.5$). Down to $T=2/3,$ the results for the $%
4\times 4$ QMC and $4\times 4$ TPSC agree to better than a few percent for $%
U<6$. One can verify from Fig. \ref{TPSCvsQMC} that at $T=0.5$ for $U<6,$
infinite-size limit TPSC and $8\times 8$ QMC results agree remarkably, the
worse disagreement being less that $10\%$ at $U=6$. This is expected from
the fact that according to the discussion of the previous section,
finite-size effects are negligible in an $8\times 8$ lattice in this
temperature range. Fig. \ref{TPSC_3d} in the following section will compare
TPSC estimates of the N\'{e}el temperature with the latest QMC calculations \cite{Staudt:2000} in $d=3$. There again, $U$ equals $3/4$ of the bandwidth
seems to be the limit of validity. We stress that TPSC is in the $N=\infty $
universality class \cite{Dare:1996} so that details may differ with the
exact result in the critical region. Nevertheless, it has been checked that
even with correlation lengths of order $10$ or more, the results are still
quite accurate.

\section{Results}

\subsection{Two dimensions}

The data that is directly extracted from the QMC calculation is the total
energy per site. The entropy extracted from this data by numerical integration,
Eq.(\ref{s_integration}), is plotted as a function of $\beta U/4=U/\left(
4T\right) $ on a logarithmic scale for different values of $U$ in Fig. \ref%
{S_QMC} along with the exact atomic (single site) limit
\begin{figure}[tbp]
\includegraphics[width=8.5cm]{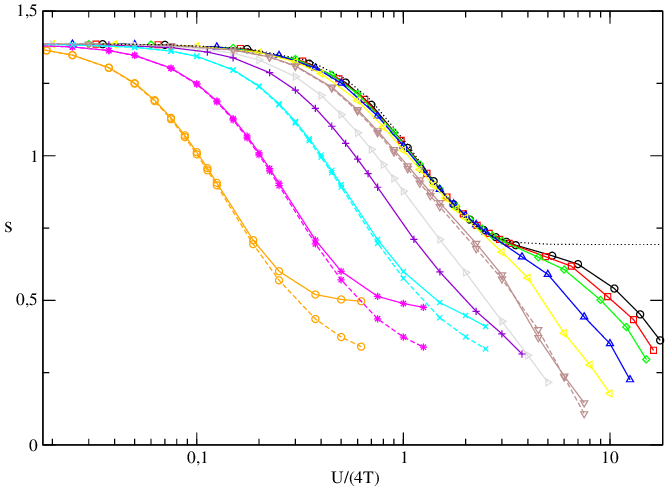}
\caption{(Color online) Entropy as a function of $\protect\beta U/4$. From
bottom to top, increasing values of $U=0.5,1,2,3,4,6,8,10,12,13,14$ are
displayed. Solid lines are for $4\times 4$ lattice and dashed lines for $%
8\times 8$ lattice. Dotted line is the exact atomic limit result for $U=14$.}
\label{S_QMC}
\end{figure}
\begin{equation}
s_{atomic}\left( \beta ,U\right) =\ln \left( 4\cosh \left( \frac{\beta U}{4}%
\right) \right) -\frac{\beta U}{4}\tanh \left( \frac{\beta U}{4}\right) .
\label{s_atomic}
\end{equation}%
Also plotted in Fig. \ref{S_QMC} are the data for an $8\times 8$ lattice
when $U=0.5,1,2,$ $6$. One can check that for $\beta <1$ $\left(
4T>W/2\right) $ and $U>W,$ where $W=8$ is the bandwidth, the above simple
formula describes the data to better than $4\%$ accuracy. This is consistent
with earlier QMC results \cite{Paiva:2001} that found that for $U=10$ the
specific heat above $T=1$ is well described by the atomic limit. Limiting
cases of the above formula are interesting. At infinite temperature, $\beta
U=0,$ one recovers the expected $\ln 4$ entropy with the first correction
given by%
\begin{equation}
s_{atomic}\left( \beta ,U\right) =\ln 4-\left( \beta U\right) ^{2}/32+\emph{O%
}\left( \left( \beta U\right) ^{4}\right) .  \label{s_atomic_beta_small}
\end{equation}%
In the $\beta U=\infty $ limit, only spin entropy is left so the atomic
limit result Eq.(\ref{s_atomic}) reduces to $\ln 2.$

At small values of the entropy, the curves in Fig. \ref{S_QMC}$\,$at small $%
U $ have a break. This can be understood as a finite-size effect given that $%
s\sim 0.5$ is the residual entropy for a $4\times 4$ lattice at $U=0.$ Lower
entropies can be reached at larger $U$ without size effects since $U$ lifts
the Fermi surface degeneracy. \cite{Note-Degenerescence} In addition, the
results for a $8\times 8$ lattice and small $U$ shown by dashed lines do
extend to lower values of the entropy.
\begin{figure}[tbp]
\includegraphics[width=8.5cm]{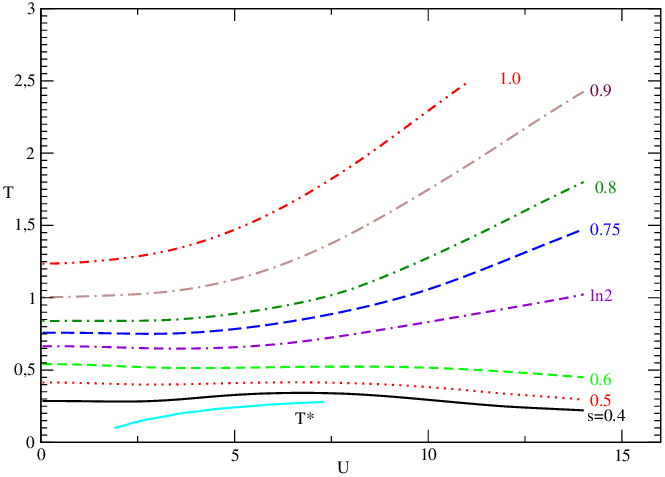}
\caption{(Color online) Isentropic curves for $d=2$ extracted from QMC
simulations, including the results of the $8\times 8$ lattice when they
differ from those of the $4\times 4$. Increasing values of $s$ are displayed
from bottom to top. The first line above the horizontal axis is the value of the crossover temperature $T^*$ determined from TPSC. It stops at strong coupling where TPSC ceases to be accurate.}
\label{isentropic-2d-QMC}
\end{figure}

Isentropic curves in Fig. \ref{isentropic-2d-QMC} are plotted in units of $T/t$ and $U/t$. They are obtained from interpolation of the QMC entropy, except for the first line above the horizontal axis that represents the value of the crossover temperature $T^*$ obtained from TPSC. \cite{tstar} As discussed
above, the data in the upper right sector $U>8,$ $T>1$ are quite accurately
explained by the atomic limit. It should be stressed that the slow variation
of entropy with $T$ and $U$ translates into inaccuracies in the
interpolation of the isentropic curves that can reach about $10\%$ in this
regime. When $U<4T$ and $W<8T$, one would expect that the high-temperature
perturbative result
\begin{equation}
s\left( \beta ,U\right) =s\left( \beta ,0\right) -\frac{1}{32}\left( \beta
U\right) ^{2}+\emph{O}\left( \left( \beta U\right) ^{4}\right) +\emph{O}%
\left( \beta ^{3}U^{2}W\right)  \label{s_perturbative}
\end{equation}%
derived in the appendices should describe well the QMC data. In fact, the
term $\emph{O}\left( \beta ^{3}U^{2}W\right) $ in the range of temperatures
shown seems to be large enough to essentially cancel the effect of the
leading $\left( \beta U\right) ^{2}$ term. The QMC isentropic curves leave
the $U=0$ axis with essentially a zero curvature and are extremely well
described by the non-interacting result $s\left( \beta ,0\right) $ in Eq.(%
\ref{s_U=0}). More specifically, for $T>1,$ $U\leq 4$ the difference between
QMC and $s\left( \beta ,0\right) $ is less than $3\%$. At $T>1$ again, the
crossover between the atomic limit value Eq.(\ref{s_atomic}) and the
non-interacting value Eq.(\ref{s_U=0}) occurs around $U=6$ where both
results differ at $T=1$ by about $10\%$ from the QMC results. That regime
does not lead to an isentropic decrease in temperature concomitant with an
increase in $U$. In the non-trivial regime where the entropy may fall with
increasing $U$ according to DMFT, \cite{Werner:2006} it is known quite
accurately that for $U=4,$ the pseudogap regime where antiferromagnetic
fluctuations are large begins around $T=0.22.$ Fig. \ref{isentropic-2d-QMC}
shows that \cite{Werner:2006} contrary to the three dimensional results of
the following section, it does not appear possible to lower the temperature
substantially by following an isentropic curve from the $U=0$ limit. Only a
small effect is observed. Even near $T=0.5,$ the isentropic curve deviates
only a little bit from the non-interacting value but it is quite close to it
up to $U$ about $6$ where a slight downturn in the isentropic curve occurs.
\cite{NoteCrossover} The flat behavior observed in this regime is confirmed
by TPSC calculations: the disagreement between the two methods down to $%
T=0.2 $ is inside the error induced by $\beta $-integration of QMC results.
The nearly horizontal isentropic is not surprising given that the minimum in
the temperature dependent double occupancy found earlier is shallow in both
TPSC \cite{Dare:2000,Lemay:2000,Kyung:2003a,Roy:2006} and QMC \cite%
{Roy:2002,Roy:2006, Brillon:2007} calculations. A very small minimum in $%
d\left( T,U=4\right) $ has been found by extrapolating double-occupancy
(local moment) to the infinite size limit in the QMC calculations of Ref. \
\cite{Paiva:2001}. Note that entry into the pseudogap (fluctuating) regime
corresponds to a rapid fall of $d$ as $T$ decreases. \cite%
{Dare:2000,Lemay:2000,Kyung:2003a,Roy:2006,Paiva:2001} In fact the
anticipation of this downfall seems to interfere with the formation of the
minimum found in higher dimension. In the regime where $d$ decreases rapidly
as temperature decreases, temperature should increase with $U$ along
isentropic curves, going in the direction opposite to the one that would be
useful for cooling from the non-interacting regime to the fluctuating phase.

From the large $U$ region, an isentropic
decrease in $U$ may also lead to a decrease in $T,$ as is obvious already from the
atomic limit result Eq.(\ref{s_atomic}). In the two-dimensional case
considered here, the results of the QMC calculation in Fig. \ref{isentropic-2d-QMC} show
that for $s<0.6$, adiabatic cooling from large $U$ is not possible. All the
temperatures along the $s=0.6$ isentropic curves are above the fluctuation
regime. Hence that regime apparently cannot be reached along a single
adiabatic curve starting from large $U$ and large $T.$ Note however the $%
s=\ln 2$ isentropic curve in Fig. \ref{isentropic-2d-QMC}. It corresponds to the high temperature spin
entropy in the large $U$ limit ($U\gg T\gg 4t^2/U$). \cite{NoteDeg} If one can trap only one of the
two atomic species per lattice site at random in the large $U$ regime, we
are in the $\ln 2$ entropy case. The lowest temperature that can be reached
by following this isentropic curve is about a factor two above the maximum
temperature where the strongly fluctuating (pseudogap) regime occurs.

\begin{figure}[tbp]
\includegraphics[width=8.5cm]{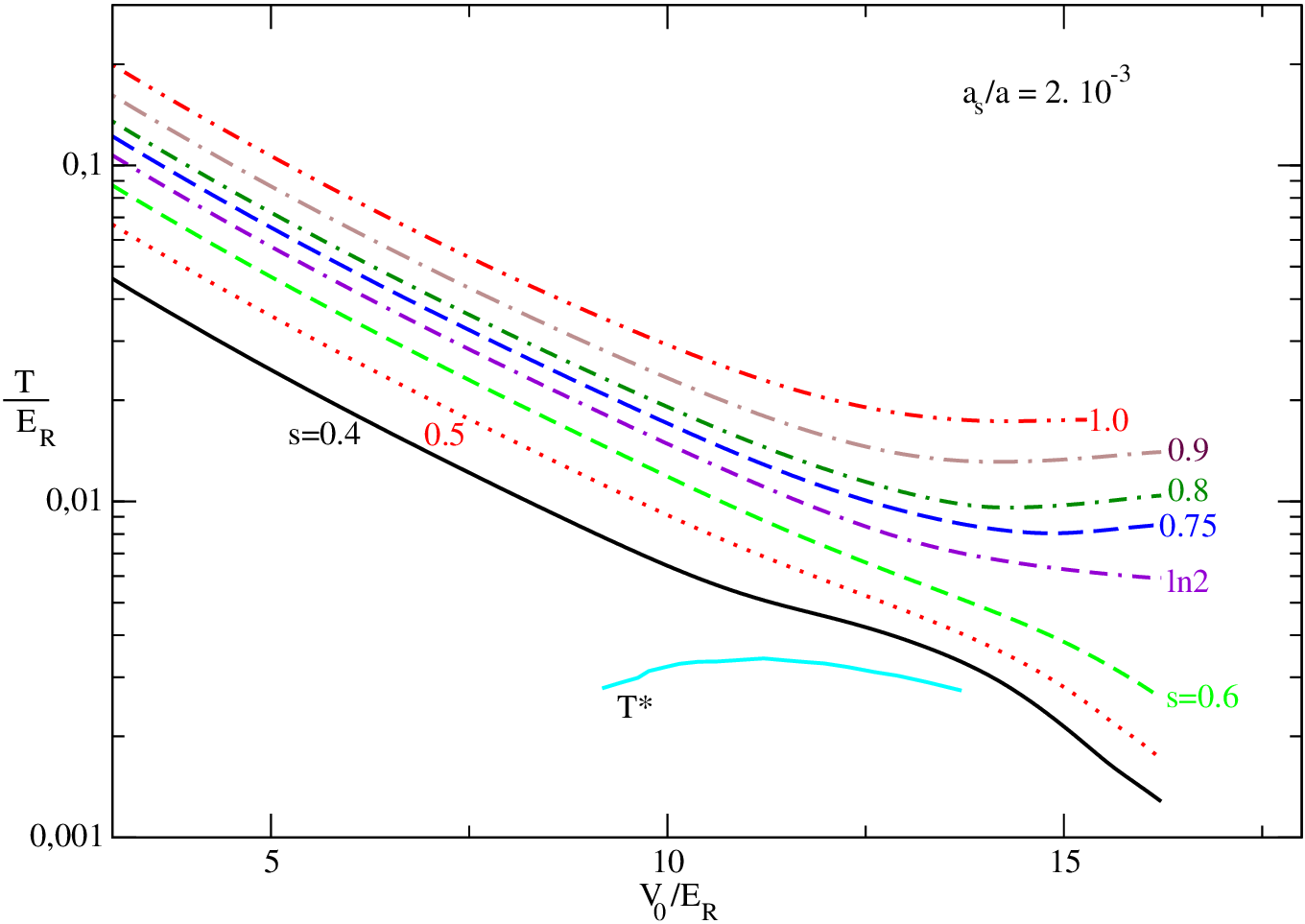}
\caption{(Color online) Isentropic curves for $d=2$ ($k_B=1$) extracted from QMC
simulations, expressed in experimental units (see text). }
\label{isent-exp-2D}
\end{figure}
We now discuss the isentropic curves in terms of
experimental parameters and units. In optical lattices, lasers create
a periodic potential defined by a period
$ a = \lambda / 2$, ( $\lambda$ is the laser wavelength) and a depth $V_0$. The energy unit conventionally used in this context is the recoil energy:
$E_R= \frac {2 \pi^2 \hbar^2 } {m \lambda^2}$, where $m$ is the mass of the fermion. To create a two-dimensional optical lattice, there is a third standing wave confining the 2D system with a depth that is large enough to prevent out-of-plane tunneling. This leads to a 2D on-site energy $U$ related to the geometric mean of the confinement strengths $U/E_R = 4 \sqrt{2 \pi} (a_s/\lambda) (V_\perp/E_R)^{1/4}(V_0/E_R)^{1/2}$ where $a_s$ is  s-wave scattering length.
\cite{Zwerger:2003,Kollath:2006}
As explained in Refs. \onlinecite{Werner:2006,Georges:2007},
there is a relation to fulfill between $a_s$, $V_0$ and
$a$ for the one-band Hubbard model to be an accurate description of cold atoms
in optical traps.

The best way to change only the interaction strength for adiabatic cooling is to change the scattering length, as can be done by tuning through a Feshbach resonance. If only the scattering length is changed, the shape of the adiabatic curves will be as in Fig. \ref{isentropic-2d-QMC}. Only the scales need to be changed. All energies in that plot are in units ($k_B=1$) of hopping $t$ which is related to recoil energy and potential strength through
$t= E_R (4 / \sqrt{\pi}) (V_0 / E_R)^{3/4} \exp{\Bigl{(} -2 \sqrt{V_0 / E_R}\Bigr{)}}$.\cite{Zwerger:2003}


We can also change $U$ by changing the potential strength $V_0$ but clearly this changes also the hopping $t$.
Thus for quantitative purposes, we also display the preceding
isentropic curves in the $(V_0/E_R,T/E_R)$
plane, rather than in the $(U/t,T/t)$ plane. This change in coordinates is discussed in Appendix \ref{Coord_2d}. It has a strong influence on the
shape of the isentropic curves as can be seen in Fig. \ref{isent-exp-2D}, obtained for the value $a_s/a = 2\ 10^{-3}$ and $U = E_R 4 \sqrt{2 \pi} (a_s / \lambda)(V_\perp / E_R)^{1/4}( V_0 / E_R)^{1/2}$ with $V_\perp / E_R = 30$. \cite{Kollath:2006} We also display in this figure the pseudogap temperature $T^*$ determined in the TPSC approach. \cite{tstar}
As $V_0 /E_R$ increases, the system cools down along isentropic curves, at least for moderate $V_0 /E_R$ values. For higher values, isentropic curves corresponding to $s> \ln{2}$ eventually bend upwards, while those corresponding to $s < \ln{2}$ bend downwards, in such a way that there is a large domain of temperatures in the vicinity of the large-repulsion spin-entropy value $s = \ln{2}$. This general behavior was also present in Fig. \ref{isentropic-2d-QMC}, and will be seen in 3D too.
We checked than the general appearance does not change for other reasonable values of
$a_s / a$ compatible with the Hubbard model.
Given that the temperature axis is displayed on a logarithmic scale, it thus appears that tuning the potential $V_0$ can be very effective in reducing the temperature of the fermions. The general cooling trend is due to the decrease in hopping $t$ associated to an increase in $V_0$. Indeed, the ratio of temperature to bandwidth is constant for isentropic curves of non-interacting electrons, hence $T$ decreases monotonically with decreasing $t$ in this case. This is the mechanism discussed in Ref.\cite{Blackie:2005}. At $V_0 < 2.3 E_R$, (not on the figure) heating can also occur. \cite{Blackie:2005} The presence of interactions can enhance the cooling compared with the non-interacting case. \cite{Werner:2006,Georges:2007}
However, cooling down the system along an isentropic by increasing $V_0$ does not necessarily mean an effective approach of the strongly fluctuating regime of the system: indeed, as can be seen from the figure, if we increase $V_0 /E_R$ further than about 12 the pseudogap region in experimental units moves away towards lower temperatures.
Note that $T^*$ determined by TPSC is not reliable at large values of $V_0 /E_R$: the limit of validity
$\frac U W \sim \frac 3 4$, corresponds to $\frac {V_0}{E_R} \sim 12$, for our choice of $a_s/a$ and $V_\perp / E_R $. \cite{Note-Texp-2D}

\subsection{Three dimensions}

\begin{figure}[tbp]
\includegraphics[width=8.5cm]{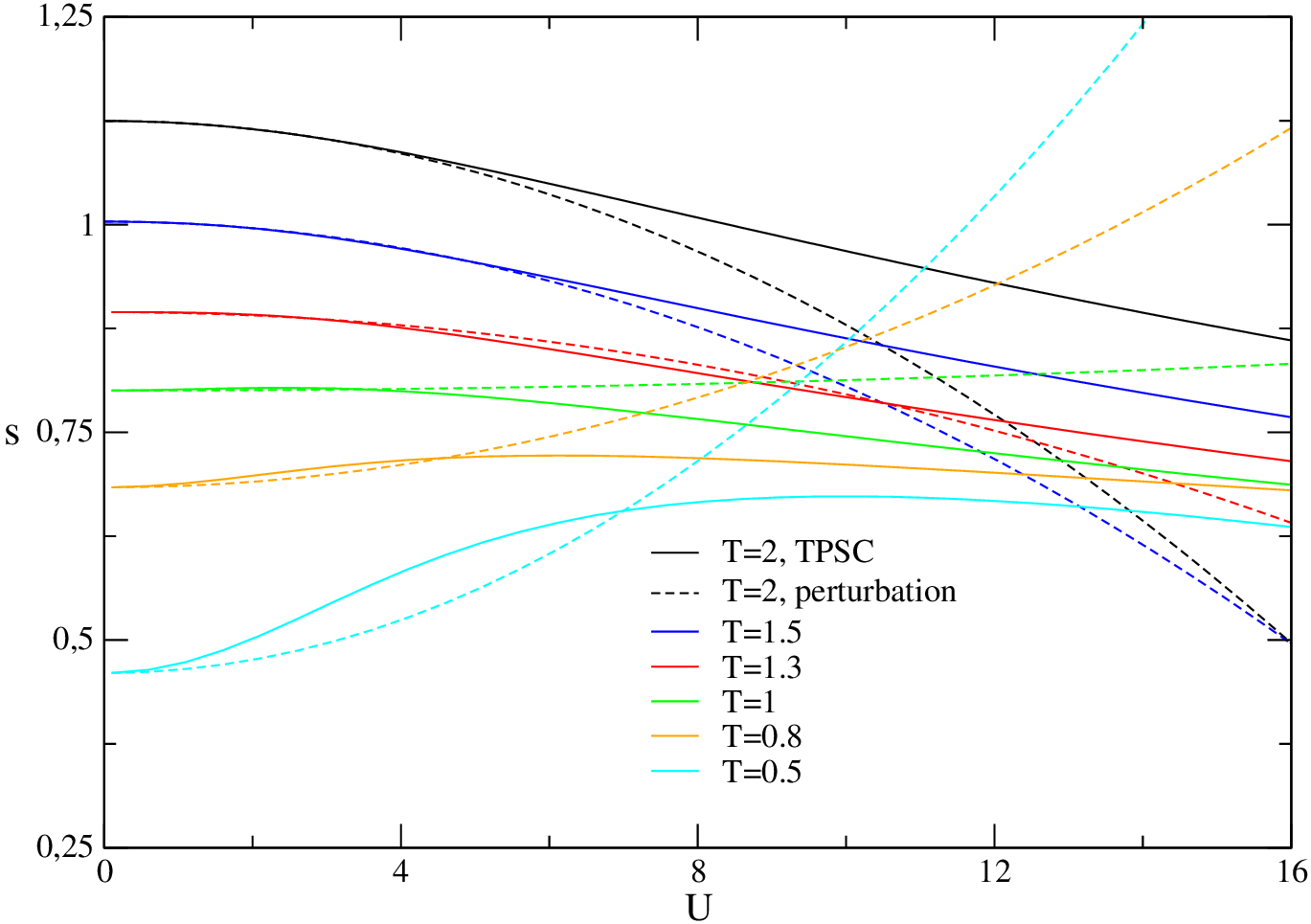}
\caption{(Color online) Comparisons of full TPSC calculation (solid lines)
with second-order perturbation theory (dashed line) for the entropy as a
function of $U$ for different temperatures. }
\label{Perturbation_2_vs_TPSC}
\end{figure}
To discuss the isentropic curves in 3D, let us go back for a while to usual units and parameters of the Hubbard model.
In three dimensions, adiabatic cooling towards the antiferromagnetic phase,
starting from small $U$, is possible and quite clearly so.
In fact, it occurs at high enough temperature that perturbation
theory (Appendix \ref{Perturbation_2_ordre}) suffices to show the effect.
This is made clear by Fig. \ref{Perturbation_2_vs_TPSC} where $\partial
s/\partial U$ changes sign from negative to positive on isothermal curves,
as $T$ decreases. By Maxwell's relation Eq.(\ref{Maxwell}), this reflects
the change in sign of $\partial d/\partial T$.  When the
temperature is large enough, the dashed lines from second
order perturbation theory agree very well, up to quite large interaction strength, with the solid lines from the full TPSC calculation

\begin{figure}[tbp]
\label{TPSC_3d}\includegraphics[width=8.5cm]{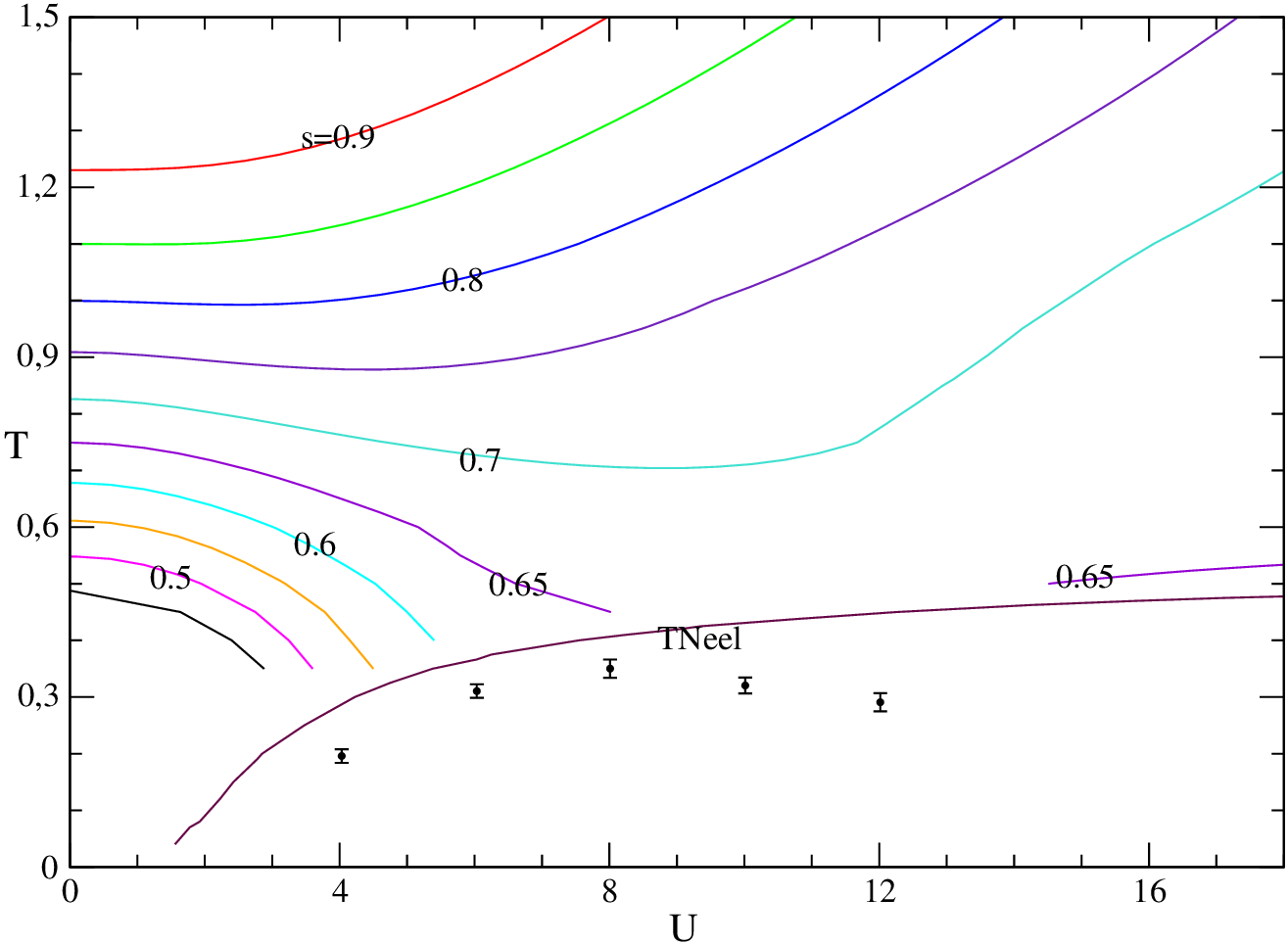}
\caption{(Color online) Isentropic curves for $d=3$ extracted from TPSC.
Increasing values of $s$ are displayed from bottom to top. The lowest solid
line is the N\'{e}el temperature. Symbols are the results of QMC
calculations taken from Ref.%
\cite{Staudt:2000}%
.}
\end{figure}

The TPSC results for the isentropic curves in three dimensions are exhibited
in Fig. \ref{TPSC_3d}. In the low-temperature regime $2\pi T/W<1,$ TPSC is
strictly valid only in the $U\ll W$ $\left( W=12\right) $ limit. This can be
checked by comparing the TPSC N\'{e}el temperature with that of the latest
QMC calculations, shown by symbols. \cite{Staudt:2000} Clearly, the agreement is satisfactory
up to $U\simeq 8$ (or $U\simeq 3W/4$ as mentioned before) where the N\'{e}el
temperature as a function of $U$ saturates according to TPSC but begins to
decrease according to QMC.

Despite the fact that TPSC is not valid in the atomic limit, it seems to
recover the correct result at high-temperature even if $U>W.$ Consider for
example $T=1.5.$ That temperature is reached along the $s=0.8$ isentropic
around $U=14$ in TPSC and around $U=13$ in DMFT. \cite{Werner:2006}
Similarly, $T=1.5$ is reached along the $s=0.75$ isentropic for $U\sim 17$
for both TPSC and DMFT. The corresponding atomic limit results, that are
dimension independent, are that $s\left( T=1.5,U=11\right) \sim 0.80$ and $%
s\left( T=1.5,U=14\right) \sim 0.75.$ TPSC is closer to DMFT than to the
high-temperature atomic limit, suggesting that both approaches take into
account the same physics at $U$ large in the high-temperature limit. Note
however that the DMFT and TPSC results are different at $U=0$ because a
model density of states is used in DMFT instead of the one following from
the exact dispersion relation used in TPSC.

As in the two-dimensional case, the value of the entropy at $T>1$ and $U<W$
is almost independent of $U.$ Contrary to the two dimensional case however,
there is a region, namely for $s\lesssim 0.65$, where cooling
along isentropic curves down to the
interesting regime is possible. Cooling however is from $T\sim 0.75$ to the maximum N%
\'{e}el temperature, $T\sim 0.4.$ DMFT predicted cooling to the N\'{e}el
temperature beginning around $T\sim 1.1.$

The possibility of adiabatically cooling all the way to the N\'{e}el
temperature starting from large $U$ is also discussed in Ref. \cite%
{Werner:2006}. For this one needs two conditions. First, the entropy at the
maximum N\'{e}el temperature has to be larger than the entropy of the
Heisenberg antiferromagnet. This condition is satisfied according to Fig. %
\ref{TPSC_3d} since the entropy at the maximum N\'{e}el temperature is
around $s_{\max }=0.65$ while the entropy of the Heisenberg antiferromagnet
is a constant $s_{H}$ estimated in Ref. \cite{Werner:2006} to be about $50\%$
smaller than $\ln 2.$ This would mean that there is indeed a maximum in the
value of the entropy at the N\'{e}el temp\'{e}rature plotted as a function
of $U.$ That maximum would be even more pronounced than that sketched in
Fig. 3 of Ref. \cite{Werner:2006}. The second condition to be satisfied is
that the temperature should decrease as $U$ decreases along isentropic
curves in the range $s_{H}<s<s_{\max }$. TPSC cannot tell whether this
condition is satisfied since for temperatures less than roughly unity
at strong coupling $U>8$ the TPSC results cannot be fully
trusted.

Incidentally, if one takes the TPSC results seriously up to $T\sim 0.7,$
then it is not possible to cool all the way to N\'{e}el temperature along
the $s=\ln 2=0.69$ \textquotedblleft infinite\textquotedblright -temperature
$\left( U/T\ll 1\right) $ isentropic curve, contrary to what DMFT suggests,
since the minimum in the $s=0.7$ isentropic curve in Fig. \ref{TPSC_3d} is
roughly a factor of two above the N\'{e}el temperature. As discussed in Ref.
\cite{Werner:2006}, DMFT does not give an accurate estimate of the entropy
at the N\'{e}el temperature since the latter is obtained in mean-field. \cite%
{NoteGeorges} This is why the $s=0.7$ isentropic curve ends at the N\'{e}el
temperature in that approximation.

\begin{figure}[tbp]
\includegraphics[width=8.5cm]{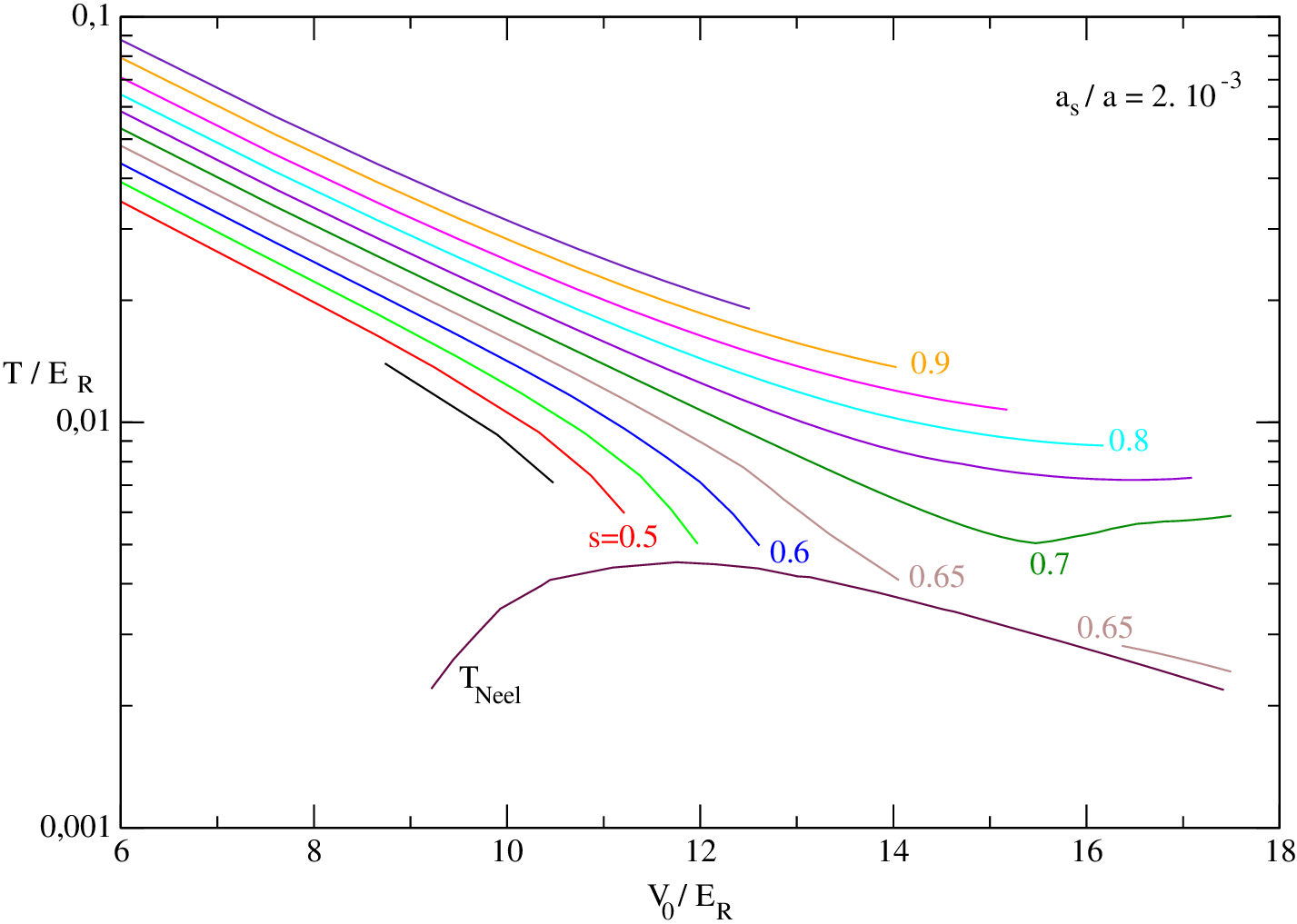}
\caption{(Color online) Isentropic curves for $d=3$ from TPSC calculations,
expressed in experimental units (see text). }
\label{isent-exp-3D}
\end{figure}

Finally, we discuss experimental units. Once again, increasing the scattering length would be the simplest way to implement directly the Pomeranchuck adiabatic cooling discussed in this paper.
Indeed varying the scattering length value will span the abscissa axis in Fig.\ref{TPSC_3d}.
All energies in that plot are in units ($k_B=1$) of hopping $t$ which is related to recoil energy and potential strength through
$t= E_R (4 / \sqrt{\pi}) (V_0 / E_R)^{3/4} \exp{\Bigl{(} -2 \sqrt{V_0 / E_R}\Bigr{)}}$.\cite{Zwerger:2003} The interaction strength on the other hand is $U = E_R 4 \sqrt{2 \pi} (a_s / \lambda)(V_0 / E_R)^{3/4}$.\cite{Zwerger:2003}

As in the 2D case, we choose to change the potential strength $V_0$, that modifies both hopping $t$ and on-site interaction $U$. This is shown in Fig.\ref{isent-exp-3D} where we display the isentropic curves and the
TPSC-determined N\'eel temperature in the experimental units \cite{Note_T_exp}. For the figure, we fix $a_s/a = 2\ 10^{-3}$. Increasing the potential $V_0$ is quite efficient to cool down the
system in absolute units, but the N\'{e}el temperature also recedes.
Two trends remain: for $s> \ln{2}$ the cooling is not sufficient to reach
the antiferromagnetic region, but a smaller entropy value may do.
In these units, the isentropic $s = \ln{2}$ seems "noisy" for large
$V_0 /E_R$ values because, in this region, the entropy surface
is flat: this makes the precise location of the isentropic curve difficult to
determine.

\section{Conclusion}

TPSC calculations confirm that the physics of adiabatic cooling by increase
of $U$ (as found in Ref. \cite%
{Werner:2006}) is correct for three dimensions. Our quantitative estimates
show a smaller but still appreciable effect. Since it occurs at relatively
high temperature, that effect is qualitatively captured already by second
order perturbation theory. Adiabatic cooling should help in reaching the N%
\'{e}el antiferromagnetic transition temperature.
Reaching the
antiferromagnetic phase should be a first step in the study of d-wave
superconductivity in optical traps. In two dimensions however, QMC shows
that this mechanism is not very effective, making it impractical to reach the
low temperature fluctuating regime by this approach.

In both $d=2$ and $d=3$ it is possible to adiabatically cool starting from
the large $U$ regime as suggested in Ref. \cite{Werner:2006}. However, in $%
d=2,$ the fluctuating regime cannot be reached along a single adiabatic
using this approach. In $d=3,$ although there are encouraging trends, we
cannot tell unambiguously with TPSC whether the N\'{e}el temperature of the
Heisenberg antiferromagnet can be reached by decreasing $U$ along a single
adiabatic that starts at high temperature.

In the context of cold fermions, changing the strength of the laser-induced lattice potential changes both hopping $t$ and interaction strength $U$. In units of absolute temperature and potential strength then, the shape of the adiabatic curves are quite different from those in the $T/t$ and $U/t$ units appropriate for the Hubbard model. While the change in $t$ can, for some cases, produce drastic cooling in absolute units, it also changes the shape of the lines for the N\'{e}el temperature (in 3D) and the pseudogap temperature (in 2D). We have plotted the results for a given scattering length as examples. Whether a given isentropic curve crosses the N\'{e}el or the pseudogap temperature is clearly independent of coordinates. To implement directly the type of interaction driven adiabatic cooling discussed in the present paper, one could change only the interaction strength by manipulating the scattering length with a Feshbach resonance.

Adiabatic cooling away from half-filling and for other cases can be studied with the methods of this paper.

\begin{acknowledgments}
We thank A. Georges, S. Hassan, R. Hayn, G. Japaridze, P. Lombardo, S. Sch\"{a}fer and J. Thywissen for useful discussions. We are especially grateful to A. Georges, B.
Kyung and J. Thywissen for a critical reading of the manuscript and for specific suggestions. Computations were performed
on the Elix2 Beowulf cluster in Sherbrooke. A.-M.S.T. would like to thank
L2MP and Universit\'{e} de Provence for their hospitality while this work
was performed. The present work was supported
by NSERC (Canada), FQRNT (Qu\'{e}bec), CFI (Canada), CIAR, and the Tier I
Canada Research Chair Program (A.-M.S.T.).
\end{acknowledgments}

\appendix%

\section{Exprimental coordinates for the two-dimensional case\label{Coord_2d}}

The problem of converting from the theoretical units $\left(  U/t,T/t\right)
$ to the experimental units $\left(  V_{0}/E_{R},T/E_{R}\right)  $ is
straightforward in three dimensions but it requires some discussion in two
dimensions. Since \cite{Zwerger:2003,Kollath:2006}
\begin{equation}
\frac{U}{E_{R}}=4\sqrt{2\pi}\left(  \frac{a_{s}}{\lambda}\right)  \left(
\frac{V_{\bot}}{E_{R}}\right)  ^{1/4}\left(  \frac{V_{0}}{E_{R}}\right)
^{1/2}%
\end{equation}
and%
\begin{equation}
\frac{t}{E_{R}}=\frac{4}{\sqrt{\pi}}\left(  \frac{V_{0}}{E_{R}}\right)
^{3/4}\exp\left(  -2\sqrt{\frac{V_{0}}{E_{R}}}\right)  .
\end{equation}
it follows immediately that%
\begin{align}
\frac{U}{t}  & =\pi\sqrt{2}\left(  \frac{a_{s}}{\lambda}\right)  \left(
\frac{V_{\bot}}{E_{R}}\right)  ^{1/4}\left(  \frac{V_{0}}{E_{R}}\right)
^{-1/4}\exp\left(  2\sqrt{\frac{V_{0}}{E_{R}}}\right)  \\
\frac{T}{t}  & =\left(  \frac{T}{E_{R}}\right)  \frac{\sqrt{\pi}}{4}\left(
\frac{V_{0}}{E_{R}}\right)  ^{-3/4}\exp\left(  2\sqrt{\frac{V_{0}}{E_{R}}%
}\right)  .
\end{align}
For definiteness we choose in this paper $a_{S}/\lambda=10^{-3}$ and $V_{\bot
}/E_{R}=30.$ Clearly, $U/t$ is not a monotonic function of $V_{0}$. For a
given $U/t$ we have two or zero real values of $V_{0}$ as we now discuss$.$

For the sake of simplicity, let us define the reduced units%
\begin{align}
u  & =\frac{U}{t}\left[  \pi\sqrt{2}\left(  \frac{a_{s}}{\lambda}\right)
\left(  \frac{V_{\bot}}{E_{R}}\right)  ^{1/4}\right]  ^{-1}\\
v_{0}  & =\sqrt{\frac{V_{0}}{E_{R}}}.
\end{align}
Then, the relation between $u$ and $v_{0}$ becomes
\begin{equation}
u=v_{0}^{-1/2}\exp\left(  2v_{0}\right)  .\label{u_v0}%
\end{equation}
This function has a minimum at $v_{0}=1/4.$ The only possible values of $u$
are thus $u\geq2\sqrt{e}$ and for each such values of $u,$ there are two
values of $v_{0},$ one less than $1/4$ and the other one larger than $1/4.$ It
is the latter that we consider as the physical value. Indeed, it corresponds
to $V_{0}/E_{R}\geq1/16$ and we know that the Hubbard model is valid only for
sufficiently large values of $V_{0}/E_{R}.$ The minimum value of $U/t$ for
$V_{\bot}/E_{R}=30$ is about $34a_{s}/\lambda$ which is quite a small value.
To solve for $v_{0}\left(  u\right)  ,$ it suffices to rewrite Eq.(\ref{u_v0})
as
\begin{equation}
-\frac{4}{u^{2}}=-4v_{0}\exp\left(  -4v_{0}\right)  .
\end{equation}
The solution to $y=x\exp\left(  x\right)  $ is the Lambert function $W_{k}$
(also known as \textquotedblleft product log\textquotedblright\ function).
Since $y\left(  x\right)  $ is non-monotonic, there exists a family $W_{k}$ of
inverse functions. If $U$ is larger or equal to the bound discussed above,
then  $0>y\geq-1/e$  and $W_{k=-1,0}$ can take real values. The branch
$x=W_{-1}\left(  y\right)  $ has  $x\leq-1$, corresponding to $V_{0}/E_{R}%
\geq1/16$ that we want to retain$.$ The other branch, $x=W_{0}\left(
y\right)  $ leads to $x\geq-1$ which we do not consider here. Hence the
solution is%
\begin{align}
\frac{V_{0}}{E_{R}}  & =\left[  -\frac{1}{4}W_{-1}\left(  -\frac{4}{u^{2}%
}\right)  \right]  ^{2}\nonumber\\
& =\left[  -\frac{1}{4}W_{-1}\left(  -\frac{t^{2}}{U^{2}}8\pi^{2}\left(
\frac{a_{s}}{\lambda}\right)  ^{2}\left(  \frac{V_{\bot}}{E_{R}}\right)
^{1/2}\right)  \right]  ^{2}%
\end{align}

\section{Entropy in the large temperature limit from the self-energy}

Consider the large Matsubara frequency (equivalently high-temperature) limit
of the self-energy, \cite{Vilk:1997}%
\begin{equation}
\Sigma \left( \mathbf{k,}ik_{n}\right) =Un_{-\sigma }+\frac{U^{2}n_{\sigma
}\left( 1-n_{-\sigma }\right) }{ik_{n}}+\ldots  \label{exact_self}
\end{equation}%
This formula is valid when $\pi T$ is larger than the frequency range over
which $\Sigma ^{R\prime \prime }$ is non-zero. In practice, since in all the
diagrams that enter the calculation of the self-energy, $k_{n}=\left(
2n+1\right) \pi T$ is compared with $\varepsilon _{\mathbf{k}}$ and there is
particle-hole symmetry at half-filling, we may expect that as soon as the
Matsubara frequencies are larger than band energies of order $\pm W/2,$
namely $\pi T>W/2,$ then the expansion may apply. This is confirmed by the
numerical results in this paper. The expansion should thus be valid for $\pi
T\gg 4$ in $d=2,$ and $\pi T\gg 6$ in $d=3$. If, inspired by the exact
atomic result Eq.(\ref{s_atomic}) we replace $\pi $ by $4$ we recover the
limits of validity mentioned in the text. In addition to this restriction on
temperature compared with bandwidth, we note that this asymptotic expansion
for the self-energy Eq. (\ref{exact_self}) is clearly a power series in $%
U/\pi T$. At half-filling, using the usual canonical transformation to the
attractive Hubbard model, we can see that if we absorb the Hartree-Fock term
in the definition of the chemical potential then only even powers of $U$
enter the expansion, which makes it convergent even faster. In addition, it
turns out that stopping the expansion of $\Sigma \left( \mathbf{k,}%
ik_{n}\right) $ at $U^{2}/ik_{n}$ at half-filling reproduces the exact
result in the atomic limit. Hence, in the special case we are interested in,
we expect that this high-temperature expansion $W/2\pi T\ll 1$ is excellent
for arbitrary values of $U,$ even if strictly speaking it should be valid
only if $U/\pi T\ll 1$ as well$.$

From the above expression for the self-energy and the sum-rule, \cite%
{Vilk:1997}
\begin{equation}
\frac{T}{N}\sum_{n}\sum_{\mathbf{k}}\Sigma \left( \mathbf{k,}ik_{n}\right)
G\left( \mathbf{k,}ik_{n}\right) e^{-ik_{n}0^{-}}=U\left\langle n_{\uparrow
}n_{\downarrow }\right\rangle  \label{Consistency}
\end{equation}%
we can extract the double occupancy $d$ that we need to compute the entropy
in the high-temperature limit. For the Green function, we again assume that $%
W/2\pi T\ll 1.$ This means that we can insert in the previous equation%
\begin{equation}
G\left( \mathbf{k,}ik_{n}\right) =\frac{1}{ik_{n}-\frac{U^{2}}{4ik_{n}}}.
\end{equation}%
This clearly neglects band effects that would contribute to order $%
UW/\left( \pi T\right) ^{2}$ to double occupancy. The normalized sum over
wave-vectors in the sum-rule Eq.(\ref{Consistency}) contributes a factor
unity while the discrete Matsubara sum can be performed exactly. One finds,%
\begin{equation}
-\frac{U}{4}\tanh \left( \frac{U}{4T}\right) =U\left( \left\langle
n_{\uparrow }n_{\downarrow }\right\rangle -\left\langle n_{\uparrow
}\right\rangle \left\langle n_{\downarrow }\right\rangle \right) .
\end{equation}%
To extract the entropy, it suffices to use Maxwell's relation Eq.(\ref%
{Maxwell}) so that%
\begin{eqnarray}
s\left( T,U\right) &=&s\left( T,0\right) -\int_{0}^{U}\frac{\partial \left( -%
\frac{1}{4}\tanh \left( \frac{U}{4T}\right) \right) }{\partial T}dU
\label{s_atomic_improved} \\
&=&s\left( T,0\right) +\ln \left( \cosh \left( \frac{U}{4T}\right) \right) -%
\frac{U}{4T}\tanh \left( \frac{U}{4T}\right) .  \notag
\end{eqnarray}%
The above expression Eq.(\ref{s_atomic_improved}) with the exact value for $%
s\left( T,0\right) $ neglects terms of order $U^{2}W/\left( \pi T\right)
^{3}.$ In practice, we found that keeping the non-interacting value of the
entropy $s\left( T,0\right) $ in the above formula does not improve the
comparison with QMC data in the region where $W/2\pi T\ll 1$ is satisfied,
whether $U$ is small or large. When we neglect all band effects compared
with temperature, then $s\left( T,0\right) $ can be replaced by $\ln 4$ and
we recover the atomic limit result Eq.(\ref{s_atomic}) that can also be
found from elementary statistical mechanics. It is the latter result that is
useful to understand the data at large values of $U.$

Expansion of $s\left( T,U\right) $ above in powers of $U/4T$ leads to the
perturbative result Eq.(\ref{s_perturbative}). One can also arrive at this
result by directly neglecting higher powers of $U/\pi T$ and $UW/\left( \pi
T\right) ^{2}$ in the self-energy and Green function,

\begin{widetext}
\begin{eqnarray}
\frac{T}{N}\sum_{n}\sum_{\mathbf{k}}\Sigma \left( \mathbf{k,}ik_{n}\right)
G\left( \mathbf{k,}ik_{n}\right) e^{-ik_{n}0^{-}} &\simeq &\frac{T}{N}%
\sum_{n}\sum_{\mathbf{k}}\left( Un_{-\sigma }+\frac{U^{2}n_{\sigma }\left(
1-n_{-\sigma }\right) }{ik_{n}}\right) \frac{1}{ik_{n}}e^{-ik_{n}0^{-}} \\
&=&U\left\langle n_{\uparrow }\right\rangle \left\langle n_{\downarrow
}\right\rangle -T\sum_{n}\frac{U^{2}}{4\left( 2n+1\right) ^{2}\left( \pi
T\right) ^{2}} \\
&=&U\left\langle n_{\uparrow }\right\rangle \left\langle n_{\downarrow
}\right\rangle -\frac{U^{2}}{16T}=U\left\langle n_{\uparrow }n_{\downarrow
}\right\rangle
\end{eqnarray}%
\end{widetext}%
so that%
\begin{eqnarray}
s\left( T,U\right) &=&s\left( T,0\right) -\int_{0}^{U}\frac{\partial \left( -%
\frac{U}{16T}\right) }{\partial T}dU  \label{Derivation_perturbation} \\
&=&s\left( T,0\right) -\frac{U^{2}}{32T^{2}}.
\end{eqnarray}%
This result, appearing in Eq.(\ref{s_perturbative}), keeps all powers in $%
W/2\pi T,$ the leading term in $U/\pi T$ and neglects $U^{2}W/\left( \pi
T\right) ^{3}$ and higher orders (the entropy is an even function of $U$ at
half-filling). It does not however assume that $U/W<1.$ It is the large
temperature here that controls the expansion. In practice we found that the
above formula does not lead to a good description of the QMC data in any
regime, even small $U$ and large $T$, unless $s\left( T,0\right) \rightarrow
\ln 4$ in the large $U$ regime. This suggests that the corrections $\emph{O}%
\left( \left( \beta U\right) ^{4}\right) +\emph{O}\left( \beta
^{3}U^{2}W\right) $ are important and in fact cancel the leading one.

Note that in all the results of this section, the dimension occurs only in
the value of $W$ and in the value of $s\left( T,0\right) .$ The atomic limit
is independent of dimension.

\section{Second order perturbation theory and TPSC for the entropy\label%
{Perturbation_2_ordre}}

In the limit $U\ll W,$ TPSC reproduces the standard perturbative expression
for double occupancy. This can be demonstrated as follows. In TPSC, double
occupancy is obtained from the following sum rule and ansatz \cite%
{Vilk:1997,Allen:2003}
\begin{eqnarray}
n-2\langle n_{\uparrow }n_{\downarrow }\rangle &=&\frac{T}{N}\sum_{q}\frac{%
\chi _{0}(q)}{1-\frac{1}{2}U_{sp}\chi _{0}(q)}  \label{Spin} \\
U_{sp} &=&U\frac{\langle n_{\uparrow }n_{\downarrow }\rangle }{\langle
n_{\uparrow }\rangle \langle n_{\downarrow }\rangle }.
\end{eqnarray}%
We used short-hand notation for wave vector and Matsubara frequency $q=(%
\mathbf{q,}iq_{n})$. Since the self-energy is constant in the first step of
TPSC, the irreducible susceptibility takes its non-interacting Lindhard
value $\chi _{0}(q).$ In a perturbation theory in $U,$ we can expand the
right-hand side of the sum rule Eq.(\ref{Spin}) and take $U_{sp}=U$ which
leads to%
\begin{eqnarray}
n-2\langle n_{\uparrow }n_{\downarrow }\rangle &=&\frac{T}{N}\sum_{q}\left(
\chi _{0}(q)+\frac{1}{2}U\chi _{0}^{2}(q)\right)  \notag \\
&=&n-2\langle n_{\uparrow }\rangle \langle n_{\downarrow }\rangle +\frac{1}{2%
}U\frac{T}{N}\sum_{q}\chi _{0}^{2}(q)
\end{eqnarray}%
or%
\begin{equation}
\langle n_{\uparrow }n_{\downarrow }\rangle -\langle n_{\uparrow }\rangle
\langle n_{\downarrow }\rangle =-\frac{1}{4}U\frac{T}{N}\sum_{q}\chi
_{0}^{2}(q)  \label{Perturbation_exact}
\end{equation}%
which shows, as expected, that double-occupancy is decreased by repulsive
interactions compared with its Hartree-Fock value. The above corresponds to
the expression obtained from direct perturbation theory for $\langle
n_{\uparrow }n_{\downarrow }\rangle -\langle n_{\uparrow }\rangle \langle
n_{\downarrow }\rangle $. The entropy can be obtained, as usual, from
integration of the Maxwell relation Eq.(\ref{Maxwell}) using the known $%
s\left( T,U=0\right) $. This is how the perturbative result in Fig. \ref%
{Perturbation_2_vs_TPSC} was obtained.

In the high-temperature limit, we recover results of the previous section,
as we now proceed to show in two different ways. TPSC at the first level of
approximation obeys the sum rule Eq.(\ref{Consistency}) that expresses a
consistency between single-particle and two-particle quantities. The
self-energy in the $U/W<1$ and large Matsubara-frequency limit has been
found in Ref. \cite{Vilk:1997}, Eq.(E.10)%
\begin{widetext}%
\begin{equation}
\Sigma \left( \mathbf{k,}ik_{n}\right) =Un_{-\sigma }+\frac{U}{ik_{n}}\left(
\frac{U_{sp}+U_{ch}}{2}n_{-\sigma }-U_{ch}n_{-\sigma }^{2}+\frac{\left(
U_{sp}-U_{ch}\right) }{2}\left\langle n_{\uparrow }n_{\downarrow
}\right\rangle \right) +\ldots  \label{Self-Asympt_TPSC}
\end{equation}%
\end{widetext}%
In the high-temperature limit $\pi T\gg W/2$ that we are interested in, the
classical (zero-frequency) contribution dominates the sum rules used to find
$U_{sp}$ and $U_{ch}$ so that $U_{sp}=U_{ch}=U$ and one recovers that the $%
1/ik_{n}$ term has the exact form $U^{2}/\left( 4ik_{n}\right) $ used in the
previous appendix$.$ We thus recover the high-temperature perturbative
result for the entropy derived there and appearing in Eq.(\ref%
{s_perturbative}).

Another way to arrive at the same result in a more transparent way that uses
only the first step of the TPSC approach $\left( U/W<1\right) $ is to work
directly with the previous perturbative result Eq.(\ref{Perturbation_exact})
and evaluate it in the high temperature limit. But we first rederive the
perturbative result Eq.(\ref{Perturbation_exact}) from Eq.(43) of Ref.\cite%
{Vilk:1997} that is valid when the correction of double occupancy from its
Hartree-Fock value is small,%
\begin{equation}
\left\langle n_{\uparrow }n_{\downarrow }\right\rangle =\left\langle
n_{\uparrow }\right\rangle \left\langle n_{\downarrow }\right\rangle \frac{1%
}{1+\Lambda U}.  \label{Double_approximate}
\end{equation}%
Correcting a factor of $2$ misprint in Ref.\cite{Vilk:1997}, the quantity $%
\Lambda $ is given by%
\begin{equation}
\Lambda =\frac{1}{n^{2}}\frac{T}{N}\sum_{iq_{n}}\sum_{\mathbf{q}}\chi
_{0}^{2}\left( \mathbf{q,}iq_{n}\right)
\end{equation}%
with $q_{n}$ a bosonic Matsubara frequency and $\chi _{0}$ the Lindhard
function. Expanding the denominator in Eq.(\ref{Double_approximate}) and
substituting $n=1$ and $U\left\langle n_{\uparrow }\right\rangle
\left\langle n_{\downarrow }\right\rangle \sim U/4,$ we do recover the
perturbative result Eq.(\ref{Perturbation_exact}).

In the limit $W/2\pi T\ll 1$, the susceptibility $\chi _{0}$ scales as $%
1/q_{n}^{2}$ which yields terms that are smaller in powers of $W/\left( 2\pi
T\right) $ than the zero-Matsubara frequency contribution. Neglecting these
finite Matsubara frequency terms, and taking the large $W/\left( 2\pi
T\right) $ limit where $f\left( \varepsilon _{\mathbf{k}}\right) \simeq
0.5\left( 1-0.5\beta \varepsilon _{\mathbf{k}}\right) ,$ we are left with%
\begin{eqnarray}
\chi _{0}\left( \mathbf{q,}0\right) &=&\frac{-2}{N}\sum_{\mathbf{q}}\frac{%
f\left( \varepsilon _{\mathbf{k}}\right) -f\left( \varepsilon _{\mathbf{k+q}%
}\right) }{\varepsilon _{\mathbf{k}}-\varepsilon _{\mathbf{k+q}}} \\
&\simeq &\frac{\beta }{2}.
\end{eqnarray}%
>From this at $n=1$ we can evaluate that $\Lambda \simeq T/\left( 2T\right)
^{2}$ so that, to leading order, the approximate double occupancy found from
Eq.(\ref{Double_approximate}) is
\begin{equation}
\left\langle n_{\uparrow }n_{\downarrow }\right\rangle \simeq \left\langle
n_{\uparrow }\right\rangle \left\langle n_{\downarrow }\right\rangle \left(
1-\frac{U}{4T}\right) =\frac{1}{4}-\frac{U}{16T}
\end{equation}%
which leads again to the high-temperature perturbative result found at the
end of the previous appendix Eq.(\ref{Derivation_perturbation}) and hence to
Eq.(\ref{s_perturbative}). Even if this time we assumed $U/W<1$ in the last
derivation (there is no Mott gap in the one-body Green functions), the fact
that the asymptotic TPSC self-energy Eq.(\ref{Self-Asympt_TPSC}) in the
high-temperature limit reduces to $U^{2}/ik_{n}$ plus corrections that
involve two more powers of $U/\pi T$ suggests (but does not prove) that the
atomic limit is also satisfied by TPSC at high temperature. In the
high-temperature limit where $s\left( T,0\right) \rightarrow \ln 4,$ the
result that we just found, Eq.(\ref{s_perturbative}), does reduce to the
first two terms of the high-temperature series of the atomic limit. A
coincidence between atomic limit and TPSC was also noted for the attractive
Hubbard model in Ref.\cite{Atomic_TPSC}.



\begin{thebibliography}{99}
\bibitem{Note-Degenerescence} Each state can be occupied in four different
ways and there are $4\left( 1+\frac{L}{2}\right) -4-2$ Fermi surface $%
\mathbf{k}$ vectors. In the last expression we take into account that the
four states at the corners are double counted in the first term and that two
of them should not be counted because of the lattice periodicity.
\bibitem{tstar} We defined $T^*$ as the temperature at which the
enhancement of the spin-susceptibility, relatively to the non-interacting value,
reaches a factor of 100.
\bibitem{NoteCrossover} The maximum crossover temperature to the pseudogap
regime where strong antiferromagnetic fluctuations occur is also expected to
occur near $U=6,$ namely at $3/4$ of the bandwidth. Indeed the maximum N\'{e}%
el temperature is located at $3/4$ of the bandwidth in $d=3.$
\bibitem{NoteDeg} We must be in the strong coupling regime, so $U\gg t$. The temperature must be larger than the exchange energy $4t^2/U$ so that the spins are independent, but less than interaction $U$ so that double occupancy is not allowed.
\bibitem{Note-Texp-2D} Another way to characterize the temperature scales, is to define the degeneracy temperature $T_F$ as half the bandwidth, namely $4t$. Note that this is different from the degeneracy temperature of the free fermions. With this definition of $T_F$, the maximum crossover temperature is located around $0.06 T_F$.
\end{thebibliography}
\end{document}